\documentclass[prb,floats,twocolumn]{revtex4}
\usepackage{graphicx}
\begin{document}

\title {Spin polaron  in the $J_1-J_2$ Heisenberg  model }

\author{I. J. Hamad$^1$, A. E. Trumper$^1$, A. E. Feiguin$^2$, and  L. O. Manuel$^1$}

\affiliation {$^1$Instituto de F\'{\i}sica Rosario (CONICET) and
Universidad Nacional de Rosario,
Boulevard 27 de Febrero 210 bis, (2000) Rosario, Argentina\\
$^2$Microsoft Research, Station Q, University of California, Santa Barbara, 
California 93106, USA}

\vspace{4in }
\date{\today}

\begin{abstract}
We have studied the validity of the spin polaron picture in the frustrated $J_1-J_2$ 
Heisenberg model. For this purpose, we have computed the hole spectral functions for 
the N\'eel, collinear, and disordered  phases of this model, 
by means of the self-consistent Born approximation and Lanczos exact diagonalization 
on finite-size clusters. We have found that the spin polaron quasiparticle excitation 
is always  well defined for the magnetically ordered N\'eel and collinear phases, 
even in the vicinity of the magnetic quantum critical points, where the local magnetization 
vanishes. As a general feature, the effect of frustration is to increase the amplitude 
of the multimagnon states that build up the spin polaron wave function, leading to 
the reduction of the quasiparticle coherence. Based on Lanczos results, we discuss 
the validity of the spin polaron picture in the  disordered phase. 
\end{abstract}
\maketitle

\section{introduction}
\begin{figure}[h]
\vspace*{0.cm}
\includegraphics*[width=0.3\textwidth]{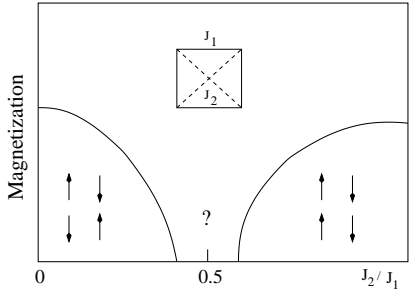}
\caption{Schematic magnetic phase diagram of the $J_1-J_2$ model. The solid lines indicates the 
values of the local magnetizations.}
\label{diagram}
\end{figure}

The antiferromagnetic (AF) $J_1-J_2$ Heisenberg Hamiltonian with competing interactions to first and 
second neighbors on the square lattice represents 
a prototypical model for the study of quantum phase transitions in low dimensional 
spin-$1/2$ Heisenberg antiferromagnets. The zero point quantum fluctuations, 
inherent of the AF Heisenberg model, can be enhanced by magnetic frustration, leading 
to an interesting ground state phase diagram which, after many years of investigations, 
has been quite well established\cite{chandra88,schulz96,trumper97,sorella98,capriotti00} and it is depicted in Fig. \ref{diagram}. As the frustration $J_2/J_1$ increases, 
the local magnetization of the N\'eel state is reduced, persisting until the quantum critical point  at about $J_2/J_1\sim 0.4$\cite{capriotti00}. 

\noindent On the other hand, for $J_2$ larger than $J_1$, the quantum collinear state 
-- neighboring spins aligned ferromagnetically along x direction and antiferromagnetically 
along y direction, or vice versa -- is stabilized by an {\it order from disorder} phenomenon\cite{villain77,chandra90}. On this side of the diagram, a decrease of $J_2/J_1$ 
implies an increase of frustration and  the zero point quantum fluctuations reduce the local 
magnetization of the collinear state which vanishes at the quantum critical point  $J_2/J_1\sim 0.6$. In the intermediate 
region $0.4 \lesssim J_2/J_1 \lesssim 0.6,$  quantum fluctuations are strong enough to destroy 
the magnetization, and the nature of the ground state is still under debate\cite{kotov00}, although the most 
sophisticated numerical method indicates that the plaquette resonance valence bound (RVB) would be 
the possible ground state\cite{capriotti00}. While the experimental realization of the quantum 
N\'eel order is widely found in many transition metal oxides, it was only recently  that the quasi-2D vanadium oxide $Li_2VOSiO_4$ was found as the first experimental realization of 
the $J_1-J_2$ model, being located at the collinear side of the phase diagram\cite{melzi00}.

A little explored issue in the literature\cite{shibata99} is the dynamics of holes injected 
in the magnetic ground states of the $J_1-J_2$ model. Since the hole strongly 
couples with the magnetic excitations, dictated by the different symmetries of the magnetic 
ground states, a strong dependence of the one-particle properties with 
frustration is expected. Hence, it is interesting to investigate whether the conventional quasiparticle 
excitations\cite{kane89}, the spin polaron picture, is valid or not in magnetically frustrated 
states. 

In the present article, we present a systematic study of the dependence of the 
one-hole spectral functions with frustration for the whole range of parameters. Using the 
self consistent Born approximation (SCBA) and exact diagonalization, our main objective 
is to investigate the validity of the spin polaron picture when the quantum critical points  
are approached from the N\'eel and the collinear sides of the phase diagram. In previous 
investigations we have already studied the hole motion in the triangular\cite{trumper04} and 
canted\cite{hamad06} antiferromagnets. We have shown that the non-collinearity of the magnetic ground states, a $120^{\circ}$ and tilted $180^{\circ}$ N\'eel  order, respectively, leads to two different mechanisms for hole motion --one magnon-assisted and other free-like -- which interfere in a non trivial way, giving rise to the vanishing of the quasiparticle weight in some regions of the Brillouin zone (BZ)\cite{trumper04,hamad06}. In the present study we have found that the spin polaron picture, which seems to be the correct description for the unfrustrated N\'eel case, remains valid for the  frustrated N\'eel and collinear phases, 
even in the vicinity of the magnetic critical points. As a general feature, the effect of frustration is to increase the contribution of multimagnon states in the spin polaron wave function leading to the reduction of the QP weight,  which always  remains finite. 
For the N\'eel case, the quasiparticle dispersion has always the same structure, with a ground state momentum at ${\bf k}=(\pi/2,\pi/2)$, even when the number of magnons increases as the disordered phase is reached. On the other hand, for the collinear phase, the structure of QP dispersion changes notably with frustration due to the interplay between the magnon assisted and free-like mechanisms for hole motion, leading to the divergence of the QP effective mass for some momenta. 
In the disordered regime, we have performed an exact diagonalization study on finite-size clusters. Guided by the SCBA, the exact spectra can be interpreted as the result of the coupling of the hole with a superposition of N\'eel and collinear short range correlations. However, it is difficult to extrapolate this result in the disordered phase since it is believed that, in the thermodynamic limit, the nature of the low lying magnetic excitations above the candidate ground state\cite{capriotti00} 
and their eventual coupling with the hole could change dramatically.

The paper is organized as follows. In Sec. II we present the self-consistent Born approximation, and we compare its predictions with exact calculations on finite-size clusters. In Sec. III we analyze the hole spectral functions obtained for the different ground states of the $J_1-J_2$ antiferromagnet. Finally, in Sec. IV we state the conclusions and discuss the validity of the spin polaron picture in the general context of frustrated antiferromagnets. 

\section{Model and Methods}
We consider the $t-J_1-J_2$ model to describe the motion of a hole in 
the frustrated $J_1-J_2$ antiferromagnet, which is written as 
\begin{eqnarray}
H & = &-t\sum_{<ij> \sigma}\left(\tilde{c}^{\dagger}_{i\sigma}\tilde{c}_{j\sigma}+
\textrm{H.c}\right)\\
\nonumber
& & +J_1 \sum_{<ij>}{\bf S}_i.{\bf S}_j+
J_2\sum_{<ik>}{\bf S}_{i}.{\bf S}_k,
\end{eqnarray}
where $<ij>$ and $<ik>$ stand for summation over first-
and second neighbors on a $N$ sites square lattice, respectively. For simplicity, 
we consider only hopping terms $t$ to first neighbors. 
The electronic operators are the projected ones, 
$\tilde{c}_{i\sigma}=(1-n_{i -\sigma})c_{i\sigma},$ that obey the no double 
occupancy constraint. 
To compute the hole spectral function of the $t-J_1-J_2$ model in the 
thermodynamic limit, we appeal to the spinless fermion representation and 
the self-consistent Born approximation\cite{martinez91,liu92}. 

\subsection{Self-consistent Born approximation}
Within the SCBA scheme, the hole is represented by a spinless fermion $h_i,$ which
carries the charge degree of freedom, while the magnetic excitations are treated 
in the spin-wave approximation (SWA), by means of the Holstein-Primakov bosons $a_i$.  
Since the SWA is valid only for magnetically ordered states, with magnons as their 
elementary excitations, we restrict the use of the SCBA to the N\'eel and collinear
phases of the $J_1-J_2$ antiferromagnet. 
Following a standard procedure\cite{trumper04}, we obtain the effective 
Hamiltonian for the hole motion
\begin{eqnarray}
\label{heff}
H_{eff}&=&\sum_{\bf k}\varepsilon_{\bf k} h^{\dagger}_{\bf k}
h_{\bf k}+\sum_{\bf q}\omega_{{\bf q}}\alpha^{\dagger}_{\bf q}\alpha_{\bf q}\\
&&+\sum_{\bf k, q}
\left(M_{\bf kq} h^{\dagger}_{\bf k}h_{\bf q} \alpha^{\dagger}_{{\bf q}-
{\bf k}}+\textrm{H.c.}\right),
\nonumber
\end{eqnarray}
where $h_{\bf k}$ is the Fourier transform of the spinless fermion operator, and 
$\alpha_{\bf q}$ are the Bogoliubov operators that diagonalize the Heisenberg part
in the SWA. 

The effective Hamiltonian comprises three terms: i) a free-like hopping term that  
takes into account the possibility of the hole to move without disturbing the 
underlying magnetic background. It is characterized by the hole dispersion 
$$\varepsilon_{{\bf k}}=\sum_{{\bf \delta}}t_{{\bf \delta}}\cos \frac{{\bf Q}.{\bf \delta}}{2}
\cos {{\bf k}}.{{\bf \delta}},$$ where $\sum_{\bf \delta}$ is the summation over neighbors ${\bf \delta}$ of 
a given site, connected by the hopping term $t_{\bf \delta}$ (first neighbors in our case). 
${\bf Q}$ is the magnetic wave vector, $(\pi,\pi)$ for the N\'eel phase, and $(\pi,0)$ 
or $(0,\pi)$ for the collinear phase. It should be noticed that a dispersive bare hole 
band will exist only for magnetic backgrounds with some non-collinear or ferromagnetic 
correlations between nearest neighbors, like in the collinear phases. Instead, in the 
N\'eel phase, $\varepsilon_{\bf k}=0$, and this fact will affect considerably the 
structure of the hole spectral function. 

\noindent ii) a free magnon term, characterized by the dispersion in the SWA
$$\omega_{\bf q}=\sqrt{\left(\gamma_{\bf q}-\Gamma\right)^2-\xi_{\bf q}^2},$$
where 
$$\gamma_{{\bf k}}=\frac{1}{2}\sum_{{\bf \delta}}J_{{\bf \delta}}\cos^2 \frac{{\bf Q}.{\bf \delta}}{2}
\cos {{\bf k}}.{{\bf \delta}},\;\;\Gamma=\frac{1}{2}\sum_{\bf \delta}J_{\bf \delta}\cos {{\bf Q}}.{{\bf \delta}},$$
and $$\xi_{\bf k}=\frac{1}{2}\sum_{\bf \delta}J_{\bf \delta}\sin^2 \frac{{\bf Q}.{\bf \delta}}{2}
\cos {{\bf k}}.{\bf \delta}.$$
iii) a hole-magnon interaction term, which incorporates the magnon-assisted 
mechanism for the hole motion, characterized by the hole-magnon 
vertex interaction
$$M_{{\bf k} {\bf q}}=\frac{i}{\sqrt{N}}\left(\beta_{{\bf q}}v_{{\bf k}-{\bf q}}-\beta_{{\bf k}}u_{{\bf q}-{\bf k}}\right),$$
where 
$$\beta_{{\bf k}}=\sum_{\bf \delta}t_{{\bf \delta}}\sin \frac{{\bf Q}.{\bf \delta}}{2}
\sin {{\bf k}}.{{\bf \delta}},$$
and the Bogoliubov coefficients are
$$u_{\bf q}=\sqrt{\frac{\gamma_{\bf q}-\Gamma+\omega_{\bf q}}{2\omega_{\bf q}}},$$
$$v_{\bf q}=\textrm{sgn}(\xi_{\bf q})\sqrt{\frac{\gamma_{\bf q}-\Gamma-\omega_{\bf q}}{2\omega_{\bf q}}}.$$

To study the hole dynamics, we compute the hole spectral function, that is, 
$A_{\bf k}(\omega)=-(1/\pi)\textrm{Im} G^h_{\bf k}(\omega)$, where 
$G^h_{\bf k}(\omega)=<AF|h_{\bf k}[1/(\omega+i\eta^+-H_{eff})]h^{\dagger}_{\bf k}|AF>$ is
the retarded hole Green function, and $|AF>$ is the undoped magnetic ground state in the
SWA. 
In the SCBA, the hole self-energy is given by the following self-consistent equation
$$\Sigma_{\bf k}(\omega)=\sum_{\bf q}\frac{|M_{{\bf k}{\bf q}}|^2}{\omega+i\eta^+-\omega_{\bf q}-
\varepsilon_{{\bf k}-{\bf q}}-\Sigma_{{\bf k}-{\bf q}}(\omega-\omega_{{\bf k}-{\bf q}})}.$$
This equation has been numerically solved for square lattices with up to $3600$ sites, with a linear mesh
of $10000$ frequences, and $\eta^+ = 0.01t$. 
Once the hole self-energy is obtained, the  QP spectral weight can be calculated as
 $z_{\bf k}=[1-\partial\Sigma_{\bf k}(\omega)/\partial \omega]^{-1}|_{E_{{\bf k}}}$, 
where the QP energy is given by the equation $E_{\bf k}=\epsilon_{\bf k}+\Sigma_{\bf k}(E_{\bf k})$.  

\subsection{Exact and SCBA comparison}
\begin{figure}[h]
\vspace*{0.cm}
\includegraphics*[width=0.45\textwidth]{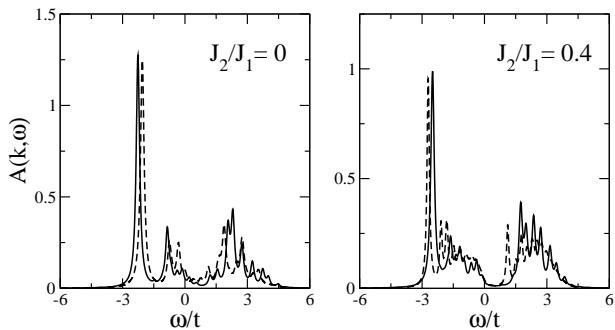}
\caption{Hole spectral functions $A_{\bf k}(\omega)$ for the N\'eel state, $N=16$,  
$J_1/t=0.4$, ${\bf k}=(\pi/2,\pi/2)$, and two different values of the frustration, $J_2/J_1=0.0$ (left) and $J_2/J_1=0.4$ (right). 
The solid and dashed lines correspond to the Lanczos and SCBA results, respectively. The Lorentzian broadening 
for Lanczos curves is $\eta^+=0.1t$}
\label{neel16}
\end{figure}

We benchmark the accuracy of the analytical SCBA by making a direct comparison of its predictions 
with Lanczos calculations performed in finite-size clusters.
In Fig. \ref{neel16} we show the 
Lanczos and SCBA hole spectral functions for a $4 \times 4$ lattice at the ground state momentum 
${\bf k}=(\pi/2,\pi/2).$ The left panel corresponds to the unfrustrated case, while  the right 
panel to the frustrated case, $J_2/J_1=0.4$. It is well known in the literature\cite{martinez91,liu92} the remarkable agreement 
between SCBA and Lanczos results for the unfrustrated case, in the strong coupling regime. This has 
been ascribed to the vanishing of the first-order vertex corrections to the SCBA hole self-energy
\cite{liu92}. From Fig. \ref{neel16}, it is evident that the good agreement remains valid in the 
frustrated case where, despite the increase of zero point quantum fluctuations, 
the SCBA reproduces the main features of the spectral function. We have 
found that SCBA yields quantitative good results for all momenta in the BZ and in the strong coupling 
limit ($J_1/t \lesssim 1$). 
\begin{table}
\caption{Exact structure factor $S^{z}_{\bf q}$ for the $J_1-J_2$ Heisenberg model, with $J_2/J_1=0.7$, $N=16$, and different values of the spatial anisotropy in the exchange interactions to first neighbors, $J_{1x}\equiv J_1$.}
\begin{tabular}{cc|cccc|cccc|cccc}
\hline
\hline

         $\bf k$  &   & &      $J_{1y}=J_{1x}$   & & && $J_{1y}=0.99J_{1x}$ &&  &  &  $J_{1y}=0.2J_{1x}$ \\
\hline
$(\pi,0)$       && &0.81590 &&&&   1.24823 &&&& 1.57717  \\
$(0,\pi)$       && &0.81590 &&&&   0.38094 &&&& 0.18420  \\
$(\pi,\pi)$       && &0.12557 &&&&   0.13307 &&&& 0.15190  \\
$(\pi/2,\pi/2)$       && &0.16624 &&&&   0.16635 &&&& 0.16667  \\
$(\pi,\pi/2)$       && &0.27813 &&&&   0.35594 &&&& 0.28422  \\
$(\pi/2,\pi)$       && &0.27813 &&&&   0.19768 &&&& 0.16669  \\
\hline
\end{tabular}
\label{table1}
\end{table}   

\begin{figure}[ht]
\vspace*{0.cm}
\includegraphics*[width=0.5\textwidth]{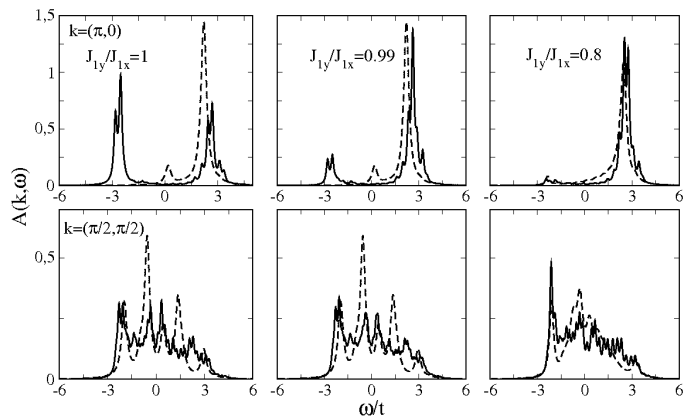}
\caption{Hole spectral function $A_{\bf k}(\omega)$ for the collinear state, $N=16$,   $J_1/t=0.4$, $J_2/J_1=0.7$, and momenta $(\pi,0)$ (upper panels) and $(\pi/2,\pi/2)$ (lower
panels). Left: $J_{1y}=J_{1x}$, middle: $J_{1y}=0.99J_{1x}$, right: $J_{1y}=0.8J_{1x}$; 
$J_{1x}\equiv J_1$. 
The solid and dashed lines correspond to the Lanczos and SCBA results, respectively.
}
\label{col16}
\end{figure}
The collinear phases break the Heisenberg Hamiltonian SU(2) spin-rotation symmetry, 
like in the N\'eel phase, and, in addition, the invariance under $\pi/2$ real space rotation\cite{chandra90}. 
This discrete two-fold degeneracy cannot be spontaneously broken in Lanczos exact calculations 
on finite-size clusters, resulting in the equal presence of $(\pi,0)$ and $(0,\pi)$ correlations 
in the magnetic background. On the other hand, in SCBA this discrete degeneracy is broken from the 
onset, with the selection of one of the two possible magnetic wave vectors, $(\pi,0)$ or $(0,\pi)$. 
Therefore, one should not naively compare Lanczos and SCBA results, until the degeneracy is somehow broken  
in Lanczos calculations, selecting one of the collinear orders. To this end, we consider spatial anisotropy in the exchange interactions to 
first neighbors, that is, different values of $J_1$ along the $x$ direction ($J_{1x}$) and the $y$ 
direction ($J_{1y}$). Clearly, if $J_{1x}$  is larger than $J_{1y}$, $(\pi,0)$ correlations will be 
favored. In fact, from Table I it results that even a slight anisotropy, $J_{1y}=0.99J_{1x}$, breaks 
the two-fold degeneracy in the collinear regime and considerably favors $(\pi,0)$ correlations.

Fig. \ref{col16} displays the Lanczos and SCBA spectral functions, for $N=16$ sites, in the 
collinear regime $J_2/J_1=0.7.$ In the isotropic case and for ${\bf k}=(\pi,0)$ (upper left panel), 
there is a strong disagreement between 
both curves; the exact spectral function exhibits a two-peak structure but the SCBA
function has only a pronounced peak at high energy. This disagreement can be understood 
if we consider that, in each method, the hole moves on different magnetic background. 
While the ${\bf Q}=(\pi,0)$ collinear phase has been chosen from the onset in the
SCBA, in the Lanczos calculations $(\pi,0)$ and $(0,\pi)$ correlations are present on equal footing (see Table I). 
With the introduction of a slight amount of anisotropy, $J_{1y}=0.99J_{1x}$, the agreement between both methods improves considerably (middle upper panel). The small 
anisotropy induces almost no changes in the SCBA because its magnetic background remains
the same. In Lanczos calculations, instead, the anisotropy produces a noticeable qualitative change, since now the $(\pi,0)$ 
correlations are largely favored over the $(0,\pi)$ ones (see Table I),
resembling the SCBA results. For larger anisotropy, 
$J_{1y}=0.8J_{1x}$ (upper right panel), the agreement between exact and SCBA spectral functions improves considerably, 
because the exact magnetic background is closer to the ${\bf Q}=(\pi,0)$ state assumed in the analytical 
calculations. On the other hand, for momenta along the BZ diagonal, like $(\pi/2,\pi/2)$ shown in the 
lower panels of Fig. \ref{col16}, the exact hole spectral functions do not depend on the anisotropy, always showing a good agreement with the SCBA spectra. 

The performed comparison lends reliability to the SCBA predictions for both frustrated phases, N\'eel and collinear. In the next section, we present SCBA results for the hole dynamics in the thermodynamic limit. 

\section{Results}

\subsection{Frustrated N\'eel case}

\begin{figure}[ht]
\vspace*{0.cm}
\includegraphics*[width=0.4\textwidth,angle=-90]{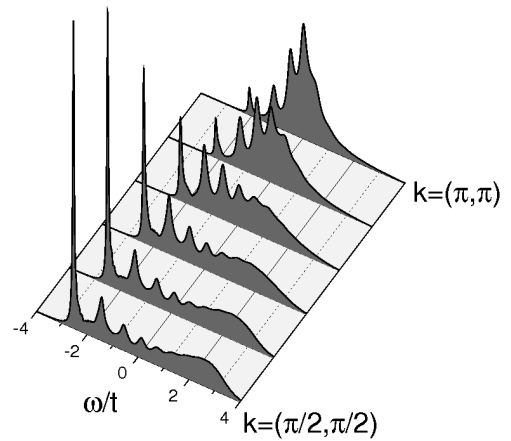}
\caption{Hole spectral function  $A_{\bf k}(\omega)$ for the unfrustrated $J_2/J_1=0$ 
N\'eel state and $J_{1}/t=0.2$ along the momentum line $(\pi/2,\pi/2)\rightarrow (\pi,\pi)$.}
\label{spec0.0}
\end{figure}

In Fig. \ref{spec0.0} we show the SCBA spectra for the unfrustrated case, $J_2/J_1=0$, along the momentum line $(\pi/2,\pi/2)\rightarrow (\pi,\pi)$ and  $J_1/t=0.2$. For this strong coupling regime, we observe a strong ${\bf k}-$dependence of the spectra due to the underlying magnetic N\'eel state. As a general feature, at low energy there is  a spin-polaron QP excitation, and several resonances above it, the so called string excitations, whose energies scale as $E_{string}\sim(J_1/t)^{2/3}$, for all momenta. This structure of the spectra has been widely confirmed by several numerical and analytical techniques\cite{liu92,dagotto94,brunner00}. While the QP excitation results from the coherent coupling of the hole motion with the AF spin fluctuations, the higher energy string excitations correspond to the hole motion inside a linear potential generated by the overturned spins. 
The existence of a coherent QP is assigned to the zero point quantum fluctuations that repair pairs of misaligned spins at a characteristic time $1/J_1$, while the hole overturns spins at a characteristic time $1/t$. Although this polaron picture is valid for the weak coupling regime $J>t$, the very existence of the spin polaron excitations in the strong coupling regime $t>J$ gives support to the idea that the coherence between hole motion and spin fluctuations is  partially preserved.

\begin{figure}[ht]
\vspace*{0.cm}
\includegraphics*[width=0.43\textwidth,angle=-90]{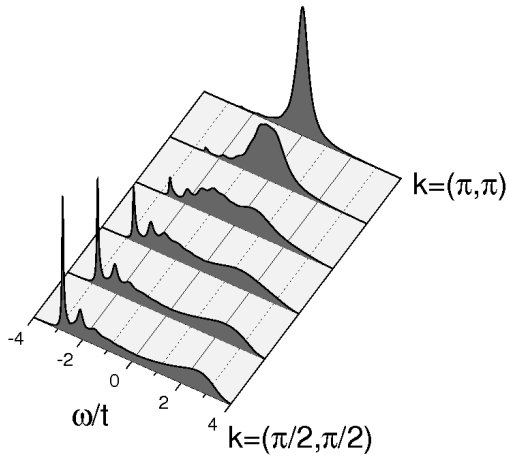}
\caption{Hole spectral function $A_{\bf k}(\omega)$ for the frustrated N\'eel state $J_2/J_1=0.3$ and $J_{1}/t=0.2$ along the momentum line $(\pi/2,\pi/2)\rightarrow (\pi,\pi)$.}
\label{spec0.3}
\end{figure}

The effect of frustration is shown in Fig. \ref{spec0.3} for $J_2/J_1=0.3$. We observe  a spectral weight reduction of both, the spin polaron and the string excitations, although the QP weight of the spin polaron remains always finite for the whole BZ, even for the extreme frustrated case (see below). Furthermore, there is a redistribution of the spectral weight at low energies which is related to the fact that the only mechanism for hole motion is magnon assisted.
At the magnetic level, frustration induces a reduction of the magnon dispersion bandwidth which can be associated with a reduced effective exchange $J_{eff}$. Therefore, the loss of QP coherence can be ascribed to an increase of the characteristic time of the spin fluctuations $1/J_{eff}$. In addition, it is also known that the spin-stiffness is smaller for the frustrated case, making the N\'eel order less rigid. Such a softening of the magnetic correlations  explains the weakening of the linear potential seen by the hole with the subsequent broadening of the string excitations. A similar effect of frustration on the  QP excitation has been observed by Shibata {\it et al.}\cite{shibata99} although their calculation, motivated by the description of the photoemission spectra of  $Sr_2CuO_2Cl_2$, was performed on an extended $t-t'-t''-J$ model.

\begin{figure}[ht]
\vspace*{0.cm}
\includegraphics*[width=0.4\textwidth]{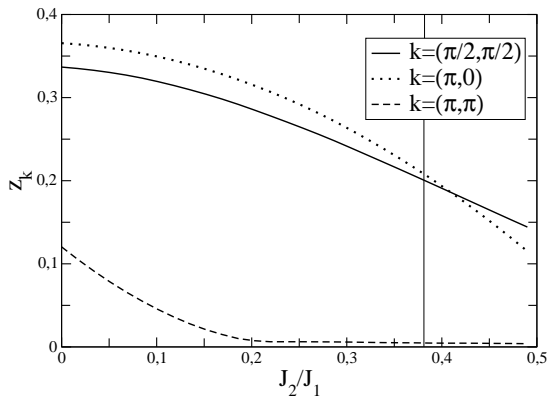}
\caption{Quasiparticle weight of the spin polaron as a function of frustration $J_2/J_1$. On the right of the vertical line  a short range N\'eel state has been assumed in the calculation.}
\label{zneel}
\end{figure}

The other issue we are concerned with, is the validity of the spin polaron picture near the magnetic quantum critical point $J_2/J_1\sim 0.4$\cite{note}. 
In Fig. \ref{zneel} we show the behavior of the QP weight $z_{\bf k}$ as a function of frustration $J_2/J_1$. It is observed a monotonic decrease of the QP weight which remains finite for all frustration and for the whole BZ. Since near the critical point the local magnetization is $m\sim 0$, the use of the spin wave theory for the calculation of the QP weight can be questioned. In order to have reliable results around the critical point, we have implemented the SCBA with a modified spin wave theory which only incorporates short range N\'eel correlations with gapped magnetic excitations. The resulting QP weight is shown on the right of the vertical line of Fig. \ref{zneel}. In particular, we confirmed that the SCBA spectra for long and short range correlations are practically the same what means that hole dynamics only depends on the short range magnetic correlations. To understand the behavior of the QP weight with frustration it is instructive to write down the spin polaron wave function as\cite{reiter94}
\begin{equation}
\label{wavefunction}
|\Psi_{\bf k}\!>=a^{(0)}_{\bf k}h^{\dagger}_{\bf k}|AF\!>+
\sum_{{\bf q'}}a^{(1)}_{\bf k\bf q'}h^{\dagger}_{\bf k\!-\!q'}{\alpha}^{\dagger}_{\bf q'}|AF\!>+\cdots,
\end{equation}
where the first term represents a state with one hole and no magnons, the second term one hole with one magnon, and so on. The coefficients $a^{(n)}$ are the probability amplitude to have $n$ magnons in the spin polaron state. As the effect of frustration is to reduce the magnon dispersion bandwidth, it is easier for the hole to emit and absorb magnons. It is then obvious that, from renormalization condition $<\Psi_{\bf k}|\Psi_{\bf k}>=1$, the increasing contribution of multimagnon states will be accompanied  by   the reduction of the QP weight $a_{\bf k}^0=z^{1/2}_{\bf k}$. Alternatively,  we can extend the approximated expression for the QP weight, derived by Kane {\it et al.}\cite{kane89}, to the frustrated case,
$$
z_{\bf k}\le \left({1+\frac{t^2 m^*z_{\bf k}}{J_1\sqrt{1-2J_2/J_1}}}\right)^{-1},
$$ 
where $m^*$ is the spin polaron effective mass, to make evident the reduction of the QP weight with increasing frustration.
\begin{figure}[h]
\vspace*{0.cm}
\includegraphics*[width=0.4\textwidth]{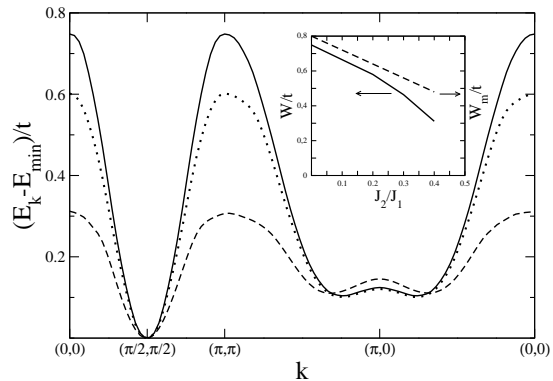}
\caption{Quasiparticle energy dispersion along the momentum line $(0,0)-(\pi,\pi)-(\pi,0)-(0,0)$, for several values of frustration
in the N\'eel state.  $J_2/J_1=0$ (solid line), $J_2/J_1=0.2$ (dotted line), $J_2/J_1=0.4$ (dashed line). Inset: bandwidth dispersion of the quasiparticle $W$ (solid line) and magnetic $W_m$ (dashed line) excitations as a function of frustration.}
\label{dispneel}
\end{figure}

Fig. \ref{dispneel} displays the spin polaron dispersion for several values of frustration. It can be seen that the reduction of the QP bandwidth $W$  is related to the reduction of the magnetic bandwidth $W_{m}=2(J_2-J_1)$, see inset of Fig. \ref{dispneel}. This is due to the fact that the only mechanism for hole motion, in the N\'eel state, is the magnon assisted one. Another consequence is the unchanged structure of the QP dispersion with frustration, which is also in accordance with the spectral weight redistribution at low energies stated above. The unchanged structure of the QP dispersion along with the reduction of the bandwidth and the QP weight, are all consistent with the intuitive picture of considering the effect of frustration 
on the hole dynamics as a loss of QP coherence  due to a reduced effective exchange $J_{eff}$.

\subsection{Frustrated Collinear case}

\begin{figure}[ht]
\vspace*{0.cm}
\includegraphics*[width=0.4\textwidth]{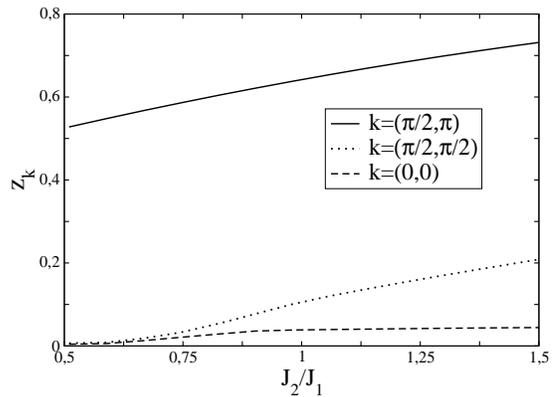}
\caption{Quasiparticle weight of the spin polaron as a function of frustration $J_2/J_1$.}
\label{zcol}
\end{figure}

\begin{figure}[ht]
\vspace*{0.cm}
\includegraphics*[width=0.4\textwidth]{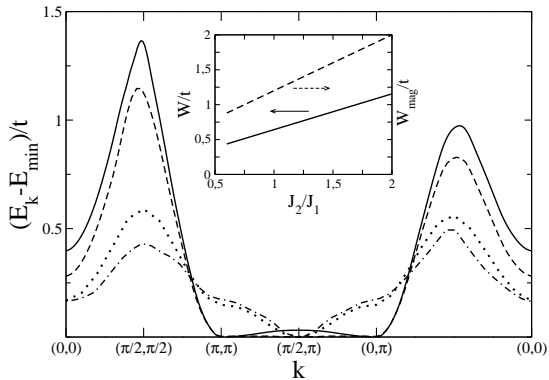}
\caption{Quasiparticle energy dispersion along the momentum line $(0,0)-(\pi,\pi)-(\pi,0)-(0,0)$, for several values of frustration
in the collinear state.  $J_2/J_1=2.5$ (solid line), $J_2/J_1=2$ (dashed line), $J_2/J_1=0.9$ (dotted line), and $J_2/J_1=0.7$ (dash-dotted). Inset: bandwidth dispersion of the quasiparticle $W$ (solid line) and magnetic $W_m$ (dashed line) excitations as a function of frustration.}
\label{dispcol}
\end{figure}

As pointed out in Section II, the hole dynamics in the collinear phase is quite different to the N\'eel phase. The arrangement  of ferromagnetic chains perpendicular to the AF chains leads to an effective Hamiltonian that involves the coexistence of the magnon assisted and free-like processes for  hole motion (see Eq. \ref{heff}).
\begin{figure}[ht]
\vspace*{0.cm}
\includegraphics*[width=0.4\textwidth,angle=-90]{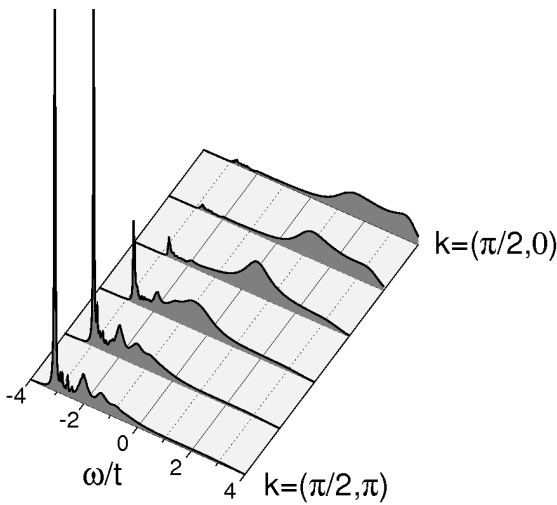}
\caption{Hole spectral function  $A_{\bf k}(\omega)$ for the collinear state ${\bf Q}=(\pi,0)$,  $J_2/J_1=1 $, and $J_{1}/t=0.2$ along the momentum line $(\pi/2,\pi)\rightarrow (\pi/2,0)$.}
\label{spec1.0}
\end{figure}
\begin{figure}[ht]
\vspace*{0.cm}
\includegraphics*[width=0.4\textwidth,angle=-90]{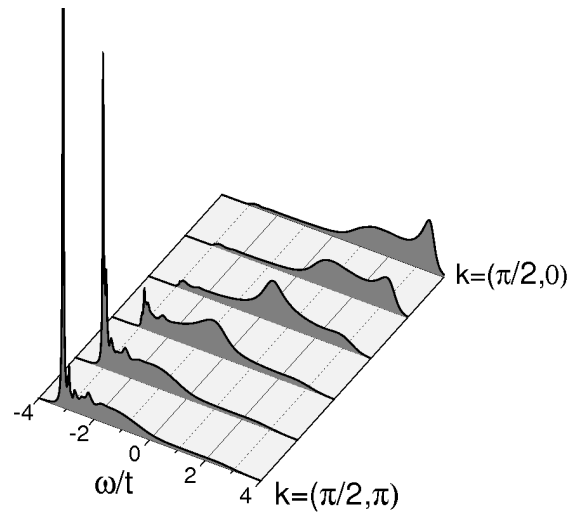}
\caption{Hole spectral function  $A_{\bf k}(\omega)$ for the more frustrated $J_2/J_1=0.6 $ collinear state ${\bf Q}=(\pi,0)$ and $J_{1}/t=0.2$ along the momentum line $(\pi/2,\pi)\rightarrow (\pi/2,0)$.}
\label{spec0.6}
\end{figure}
The presence of such processes can be identified in the hole spectral functions. In Fig. \ref{spec1.0} we show the SCBA prediction for the collinear phase ${\bf Q}=(\pi,0)$ at $J_2/J_1=1$ and $J_1/t=0.4$, along the momentum line $(\pi/2,\pi)\rightarrow (\pi/2,0)$.  For this strong coupling regime, there is again a strong ${\bf k}-$dependence of the spectra due to the underlying magnetic collinear state. The low energy part is related to the spin polaron picture where the hole is mainly coupled to the AF magnons, while at high energy there is a resonance in the spectra that is related to the hole motion along the ferromagnetic chains whose dispersion is of the tight binding form. Although the QP excitation is  originated by the AF fluctuations, the string excitations practically disappear owing to the fact that the only retraceable paths generating a linear potential are one dimensional in the collinear state . On the other hand, the 2D AF character of the N\'eel state increases noticeably the number of such retraceable paths, making the string excitations observable.
Besides the reduction of the magnetic bandwidth, another effect of frustration in the collinear state is to tune  the interference between both processes for hole motion, resulting  in a spectral weight transfer from low to high energies. See for instance, in Fig. \ref{spec0.6}, the merging shoulders at $\omega \sim 3t$ of the last three spectra, near ${\bf k}=(\pi/2,0)$,  for the more frustrated case $J_2/J_1=0.6$. Although not easily discernible, the spin polaron excitation is well defined. In particular,  Fig. \ref{zcol} shows the monotonic reduction of the QP weight as the critical point $J_2/J_1\sim 0.51$ is approached\cite{note}.
We again confirm that the spectra practically do not change when a short range collinear order with gapped excitations is assumed in the calculation near the critical point. Similarly to the   N\'eel case, the spin polaron picture is valid for the collinear regime, being the QP weight always finite for the whole BZ. We again interpret the reduction of $z_{\bf k}$ as a consequence of the increasing contribution of multimagnon terms in the spin polaron wave function of Eq. \ref{wavefunction}.
For increasing frustration --smaller $J_2/J_1$-- there is a linear reduction of the QP dispersion bandwidth, which follows the magnetic one $W_m=2J_2+J_1$ (see inset of Fig. \ref{dispcol}), and a notable change of the structure of the QP dispersion (see Fig. \ref{dispcol}). In particular, around $J_2/J_1\sim 2$ the QP minimum moves from $(\pi,\pi)$ and $(0,\pi),$ to $ (\pi/2,\pi)$ leading to the divergence of the effective mass for some momenta. 
This behavior is in contrast to that found in the N\'eel state, where the only effect of frustration was a redistribution of spectral weight at low energies. In the collinear state, instead, the spectral weight transfer is between the low and the high energy parts of the spectra which, at the microscopic level, can be ascribed to the subtle interference between the magnon assisted and free like processes for hole motion. It is by this mechanism, ultimately tuned by frustration, that the QP dispersion changes its structure. 
It is interesting to note that the same behavior -- namely, the change on the structure of the QP dispersion along with the non-vanishing of QP weight in the vicinity of the quantum critical point --  has been recently observed in the bilayer Heisenberg antiferromagnet, where the interlayer coupling $J_{\perp}$ induces a quantum disordered phase transition\cite{brunger06}.

\subsection{Disordered phase}

As pointed out in the introduction, the nature of the disordered magnetic ground state in the intermediate regime $0.4\lesssim J_2/J_1 \lesssim 0.6$ is still under debate, although there is numerical evidence of the presence of plaquette RVB correlations\cite{capriotti00}. While in the ordered regimes the spin wave excitations are the proper magnetic degrees of freedom that couple with the hole, 
the lack of a correct analytical description of the magnetic excitations above  the disordered phase  does not allow one to envisage the proper effective Hamiltonian for the hole propagation. Then, the hole dynamics in the disordered phase is not  analytically tractable within our scheme of calculation. Consequently, we resort to the Lanczos method to study the hole spectral functions on finite-size clusters. For these small systems, there is an important reduction of the disordered window relative to the thermodynamic limit. In particular, for $N=20$ the weakening of the N\'eel and the collinear correlations occurs in a narrow window around $J_2/J_1\sim 0.58$, with an enhancement of translational symmetry breaking correlations, such as  columnar and plaquette-like\cite{dagotto89}.
\begin{figure}[ht]
\vspace*{0.cm}
\includegraphics*[width=0.4\textwidth]{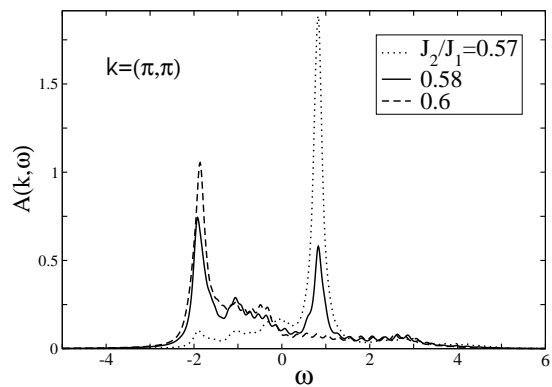}
\caption{Exact hole spectral function $A_{\bf k}(\omega)$  for a cluster  of $N=20$ sites evaluated at  ${\bf k}=(\pi,\pi)$ for the N\'eel (dotted line), collinear (dashed line), and disordered (solid line) regimes.}
\label{disordered}
\end{figure}
In Fig. \ref{disordered} we show the hole spectral functions for $N=20$ sites, momentum ${\bf k}=(\pi,\pi)$, and  three values of frustration ranging from the N\'eel ($J_2/J_1=0.57$) to the collinear ($J_2/J_1=0.6$), across the disordered regime ($J_2/J_1=0.58$). We observe a noticeable spectral weight transfer from high to low energies despite the very small change in frustration, $\Delta J_2=0.03 J_1$. This behavior can be interpreted by recalling the different mechanisms for hole motion in the N\'eel and collinear regimes. The structure of the hole spectral function for $J_2/J_1=0.57$ $(0.6)$ closely resembles  that of the magnetically ordered N\'eel (collinear) phase, in the thermodynamic limit. Surprisingly, in the disordered regime,  the hole spectral function (solid line in Fig. \ref{disordered}) seems to be the superposition of the N\'eel and collinear cases. In particular, as frustration increases the position of the peaks remains unaltered  while a spectral weight transfer takes place. Extra features in the hole spectral functions, that could indicate the coupling of the hole to translational symmetry breaking correlations, are not  observed in the disordered phase.  It is worth noticing, however, that our results correspond to a small cluster size, $N=20$, and therefore they cannot be naively extrapolated to the thermodynamic limit. Hence, the hole dynamics in the disordered phase of the $J_1-J_2$ Heisenberg model, and  the validity of the spin polaron picture in particular, remains an open problem.

\section{Concluding remarks}

In this work we have studied the dynamics of a hole injected in the magnetic ground 
states of the frustrated $J_1-J_2$ Heisenberg model. To accomplish this task we have 
used the self consistent Born approximation and exact diagonalization on finite 
clusters. Our main result is that the spin polaron scenario remains valid  whenever 
the ground state is magnetically ordered, even in the vicinity of the N\'eel and 
collinear quantum critical points. That is, there is always a well defined quasiparticle 
excitation associated with a hole dressed by magnons. Furthermore, the narrowing 
of the magnon dispersion, induced by frustration, leads to a loss of QP coherence 
due to the proliferation of multimagnon processes that build up the spin polaron 
wave function.

At this point, it is interesting to discuss the physical scenario behind the spin polaron picture in 
the general context of 2D antiferromagnets. The main idea  is based on the dominant 
pole approximation\cite{kane89} where the existence of a QP pole crucially depends
on the density of states of the low lying energy excitations which can couple to 
the hole. For the simpler N\'eel case, there are few low lying spin excitations, 
and there is very little phase space available for the hole to scatter, so the 
low energy states of the hole can have a long lifetime. In this case the 
underlying microscopic process for hole motion is the magnon assisted one. 
For the collinear state, the presence of ferromagnetic correlations induces 
free-like hole hopping processes that interfere with the magnon assisted ones. 
However the QP excitations are  still well defined even near 
the magnetic critical point. For non collinear cases, like the triangular and canted antiferromagnets, we have recently shown that the effect of such interference 
can be dramatic. In particular, for the triangular AF with  positive $t$ we have found\cite{trumper04}
 a vanishing of the QP excitation
in an ample region of the Brilouin zone. For canted N\'eel phases, we were 
able to investigate  such interference by continually tuning the canting angle, 
finding a critical value at  which the QP excitations 
disappear\cite{hamad06}. Therefore, from the present and previous studies, we argue 
that the collinearity of the N\'eel and the collinear phases are crucial for the 
existence of QP excitations. This idea is also supported by results on the honeycomb\cite{luscher06} and bilayer square lattices\cite{vojta98}.
One interesting question is the breakdown of the spin polaron excitation and the possible realization of spin-charge separation\cite{anderson06}. It was suggested that this scenario could take place in disordered magnetic states, although the lack of reliable methods to compute hole spectral functions in the thermodynamic limit makes this task quite difficult. Here the nature of the low lying magnetic excitations is crucial. For gapped magnetic excitations, like in ladder\cite{brunner01} and 
checkerboard\cite{lauchli04} systems, a coherent QP motion 
of an injected hole is expected, at least in part of the BZ. On the other hand, for the kagom\'e lattice there is a quite small triplet gap $\Delta\sim 0.05J$ and an exponentially large number of singlet excitations within such a triplet gap\cite{waldtmann98}. These unconventional low lying excitations may prevent the appearance of spin polaron excitations, as recently observed in numerical investigations\cite{lauchli04}.
In the present work we have numerically obtained the hole spectral functions for the disordered phase of the $J_1-J_2$ model, for  $N=20$ sites. We have found that these spectral functions resemble a superposition of the N\'eel and collinear cases, suggesting the existence of quasiparticle. Although our results cannot be naively extrapolated to the thermodynamic limit, we believe that if the ground state of the disordered regime is a plaquette RVB with triplet gapped excitations, the spin polaron picture could still be  valid.

From the experimental point of view, the quasi-2D vanadium oxide $Li_2VOSiO_4$ is considered a good realization of the $J_1-J_2$ model, located at the collinear side of the diagram\cite{melzi00}. Recent NMR experiments\cite{carretta02} revealed a very low spin dynamics which has been interpreted as the coexistence of $(\pi,0)$ and $(0,\pi)$ magnetic domains of collinear phases in the regime $T_N<T<J_2+J_1$. Beyond the theoretical interest in the validity of the spin polaron picture, the interplay we have established between the microscopic mechanisms and the features of the spectral functions can help to interpret experimental studies\cite{hamad07}.


\acknowledgements 
We acknowledge useful discussions with A. L. Chernyshev. This work was supported by the ANPCYT under grants PICT2004 Nro. 25724 and PICT2005 Nro. 38045. IJH thanks Fundaci\'on J. Prats for partial 
support.


\begin{thebibliography}{99}
\bibitem{note} There is some uncertainty about the exact location of the N\'eel and collinear quantum critical points. Numerical studies suggest the values 0.4 and 0.6, respectively. On the other hand, the SWA yields the values 0.38 and 0.51, respectively.
\end{thebibliography}
\end{document}